\documentstyle[preprint,aps,epsf]{revtex}

\begin{document}                                                                     
\baselineskip 22pt                                                                   
\vspace*{-0.8cm}                                                                      
\noindent
\hspace*{11.6cm}
KEK-TH-473\\
\noindent
\hspace*{11.3cm}
SNUTP 96-025\\
\noindent
\hspace*{11.5cm}
YUMS 96-005\\
\noindent
\hspace*{10.7cm}
(Modified version)\\

\vspace{-1.3cm}

\begin{center}                                                                                                                                  
{\Large \bf Average Kinetic Energy of Heavy Quark in Semileptonic $B$ Decay\\}

\vspace{0.4cm}
 
Dae Sung Hwang$^1$, ~C. S. Kim$^{2,3}$ ~and~ Wuk Namgung$^4$\\
\end{center}

\noindent$1$: Dept. of Physics, Sejong Univ. 
Seoul 143, Korea (dshwang@phy.sejong.ac.kr)\\
\noindent$2$: Dept. of Physics, Yonsei Univ. 
Seoul 120, Korea (kim@cskim.yonsei.ac.kr)\\
\noindent$3$: Theory Division, KEK, Tsukuba, 
Ibaraki 305, Japan (cskim@kekvax.kek.jp)\\
\noindent$4$: Dept. of Physics, Dongguk Univ.
Seoul 100, Korea (ngw@cakra.dongguk.ac.kr)\\

\vspace{-0.2cm}
\begin{center}                                                                                                                                  

{\bf Abstract} \\

\end{center}

Within the ACCMM model the average kinetic energy of heavy quark in 
a heavy-light meson
is calculated as $\langle {\bf p}^2 \rangle = {3 \over 2} {p_{_F}}^2$,
solely from the fact that
the Gaussian momentum probability distribution has been taken in the 
ACCMM model. 
Therefore,  the Fermi momentum parameter $p_{_F}$ of the ACCMM model is
not a truly free parameter, but is closely related to the average kinetic
energy of heavy quark, which is theoretically calculable in principle.
In this context, we determine $p_{_F}$ by comparing
the theoretical prediction
of the ACCMM model with the model independent lepton energy spectrum of
$B \rightarrow e \nu X$ from the recent CLEO analysis, and find that
$p_{_F}=0.54~\pm~^{0.16}_{0.15}$ GeV.
We also calculate $p_{_F}$ in the relativistic quark model by applying
the quantum mechanical variational method, and obtained
$p_{_F}=0.5\sim 0.6$ GeV. We show the correspondences between
the relativistic quark model and the heavy quark effective theory.
We then clarify the importance of the value of $p_{_F}$ in the
determination of $|V_{ub}/V_{cb}|$.
\\




\pagebreak 

\baselineskip 22pt

\noindent
{\bf \large 1. Introduction}\\

In the standard $SU(2) \times U(1)$ gauge theory of Glashow, Salam and
Weinberg the fermion masses and hadronic flavor changing weak transitions
have a somewhat less secure role, since they require  a prior knowledge
of the mass generation mechanism. The simplest possibility to give mass
to the fermions in the theory makes use of Yukawa interactions involving
the doublet Higgs field. These interactions give rise to the
Cabibbo--Kobayashi--Maskawa (CKM) matrix: Quarks of different flavor are
mixed in the charged weak currents by means of an unitary matrix $V$.
However, both the electromagnetic current and the weak neutral current remain
flavor diagonal. Second order weak processes such as mixing and CP--violation
are even less secure theoretically, since they can be affected by both
beyond the Standard Model virtual contributions and new physics
direct contributions. Our present understanding of CP--violation is based
on the three--family Kobayashi--Maskawa model \cite{ckm} of quarks, 
some of whose charged--current couplings have phases. Over the past decade,
new data have allowed one to refine our knowledge about parameters of 
this matrix $V$.

In the minimal Standard Model CP--violation is possible through the CKM mixing
matrix of three families, and it is important to know whether the element 
$V_{ub}$ is non-zero or not accurately. Its knowledge is also necessary to 
check whether the unitarity 
triangle  is closed or not \cite{quinn}. 
However, its experimental value is very poorly known presently and its better 
experimental information is urgently required. At present, the only 
experimental method to measure $V_{ub}$ is through the end-point lepton
energy spectrum of the inclusive $B$-meson
semileptonic decays, {\it e.g.} CLEO \cite{cleo} and ARGUS \cite{argus},
and their data indicate that $V_{ub}$ is non-zero.
Recently it has also been  suggested  that the measurements of
hadronic invariant mass spectrum \cite{kim,cskim} as well as 
hadronic energy spectrum \cite{bouzas}
in the inclusive $B \rightarrow X_{c(u)} l \nu$ decays can be
useful in extracting $|V_{ub}|$ with better theoretical understandings.
In future asymmetric $B$ factories with vertex detector, 
the hadronic invariant mass spectrum will offer
alternative ways to select $b \rightarrow u$ transitions that are much more
efficient than selecting the upper end of the lepton energy spectrum, with
much less theoretical uncertainties.

The simplest model for the semileptonic $B$-decay is the spectator model which 
considers the decaying $b$-quark in the $B$-meson as a free particle. 
The spectator model is usually used with the inclusion of perturbative QCD 
radiative corrections \cite{kuhn}.
Then the decay width of the process $B\rightarrow X_ql\nu$ is given by
\begin{eqnarray}
{\Gamma}_B (B &\rightarrow& X_ql\nu) 
\equiv |V_{qb}|^2 \times
{\widetilde{\Gamma}_B} (B\rightarrow X_ql\nu) \nonumber \\
&\simeq& {\Gamma}_b (b\rightarrow ql\nu ) =
\vert V_{qb}{\vert}^2\left({{G_F^2m_b^5}\over {192{\pi}^3}}\right)
f\left(z={{m_q}\over {m_b}}\right)
\left[1-{{2}\over {3}}{{{\alpha}_s}\over {\pi}}
g\left(z={{m_q}\over {m_b}}\right)\right]~,
\label{f1}
\end{eqnarray}
where $m_q$ is the mass of the final $q$-quark decayed from $b$-quark.
Here $f(z)$ is the phase-space factor, and $g(z)=(\pi^2-31/4)(1-z)^2+1.5$ 
is the corresponding single gluon QCD correction \cite{kimartin}.
As can be seen, the decay width of the spectator model depends on $m_b^5$, 
therefore small difference of $m_b$ would change the decay width significantly.
The model of Altarelli $et$ $al.$ \cite{alta} (ACCMM model) is an improvement
on the naive free-quark decay spectator model, but at the cost of
introducing  several free parameters: the final (charm) quark mass $m_c$,
the spectator mass $m_{sp}$, and the most important Fermi momentum
function $\phi({\bf p};p_{_F})$ that includes both binding and final state
interaction effects.

In Section 2, we determine the Fermi momentum parameter $p_{_F}$ 
by comparing the theoretical prediction
of the ACCMM model with the model independent lepton energy spectrum of
$B \rightarrow X_c l \nu$ for the whole region of electron energy, which
has been recently extracted by CLEO \cite{cleo-exp}.
Previously, the comparison had been hampered by the cascade decay of
$b \rightarrow c \rightarrow s l \nu$, and only the part of lepton energy
spectrum ($E_l > 1.8$ GeV) could be compared to give $p_{_F} \sim 0.3$ GeV.
However, we argue that the value $p_{_F} \sim 0.3$ GeV, which has been commonly 
used in experimental analyses, has no theoretical or experimental clear
justification.
Therefore, it is strongly recommended to determine the value of $p_{_F}$ more
reliably and independently, 
when we think of the importance of its role in experimental analyses.
A better determination of $p_{_F}$ is also interesting theoretically 
since it has its own physical correspondence related to the 
average kinetic energy ($\langle {\bf p}^2 \rangle$) of heavy quark 
inside heavy meson. In this context we calculate theoretically 
the value of $\langle {\bf p}^2 \rangle$
in the relativistic quark model using quantum mechanical variational 
method in Section 3.
We also compare our model with the heavy quark effective theory (HQET)
in expansion of $1/M_Q$.
The value of $p_{_F}$ is particularly important
in the determination of the value of
$|V_{ub}/V_{cb}|$, as we explain in Section 4.
Section 5 contains our conclusions.
\\

\noindent
{\bf \large 2. Determination  of $p_{_F}$ from the Experimental
Spectrum}\\

Altarelli $et$ $al.$ \cite{alta} proposed for the inclusive $B$-meson 
semileptonic decays their ACCMM model, which incorporates the bound state 
effect by treating the $b$-quark as a virtual state particle, thus giving 
momentum dependence to the $b$-quark mass.
The virtual state $b$-quark mass 
$W$ is given by
\begin{equation}
W^2({\bf p})=m_{_B}^2+m_{sp}^2-2m_{_B}{\sqrt{{\bf p}^2+m_{sp}^2}}
\label{f2}
\end{equation}
in the $B$-meson rest frame, where $m_{sp}$ is the spectator quark mass,  
$m_{_B}$ is the $B$-meson mass, and {\bf p} is the momentum of 
the $b$-quark inside $B$-meson.

For the momentum distribution of the virtual $b$-quark, Altarelli $et$ $al.$ 
considered the Fermi motion inside the $B$-meson with the Gaussian momentum
probability distribution
\begin{equation}
\phi ({\bf p};p_{_F})={{4}\over {{\sqrt{\pi}}p_{_F}^3}}e^{-{\bf p}^2/p_{_F}^2}~,
\label{f3}
\end{equation}
where the Gaussian width, $p_{_F}$, is treated as a free parameter.
Then the lepton energy spectrum of the $B$-meson  decay is given by
\begin{equation}
{{d{\Gamma}_B}\over {dE_l}}(p_{_F}, m_{sp}, m_q, m_{_B})=
{\int}_0^{p_{max}} p^2\, dp\ \phi ({\bf p};p_{_F})\ 
{{d{\Gamma}_b}\over{dE_l}}(m_b=W, m_q)~,
\label{f4}
\end{equation}
where $p_{max}$ is the maximum kinematically allowed value of $p=|{\bf p}|$.
The ACCMM model, therefore, introduces a new parameter $p_{_F}$ for the 
Gaussian momentum distribution of the $b$-quark
inside $B$-meson, instead of the $b$-quark mass of the spectator model.
In this way the ACCMM model incorporates the bound state effects and reduces
the strong dependence on $b$-quark mass in the decay width of the 
spectator model.

The Fermi momentum parameter $p_{_F}$ is the most essential parameter of  
the ACCMM model, as we explained in the above.
However, the experimental determination of its value 
from the lepton energy spectrum has been very ambiguous, because
various parameters of the ACCMM model, such as $p_{_F}$, $m_q$ and $m_{sp}$, 
are fitted all together from the limited region of end-point lepton 
energy spectrum ($E_l > 1.8$ GeV) to avoid the cascade decay of
$b \rightarrow c \rightarrow s l \nu$,
and because the perturbative QCD corrections are very sensitive
in the end-point region of the spectrum. Recently, CLEO \cite{cleo-exp}
extracted the model independent lepton energy spectrum of 
$B \rightarrow X_c l \nu$ for the whole region of electron energy
from 2.06 fb$^{-1}$ of $\Upsilon(4S)$ data, which is shown in Fig. 1, 
with much smaller uncertainties compared to the previously measured
results of ARGUS \cite{argus2}.
Now we compare the whole region of experimental electron energy spectrum 
of CLEO with the theoretical prediction of the ACCMM model, Eq. (\ref{f4}), 
to derive the value of $p_{_F}$ using $\chi^2$ analysis.
With $p_{_F}$, $m_c$ and $m_{sp}$ as free parameters, 
for one $\sigma$ standard deviation we obtain
\begin{equation}
p_{_F}~ =~ 0.54~ \pm~ ^{0.16}_{0.15}~~{\rm GeV}~,
\label{tz2}
\end{equation}

In Table I, we show the extracted values of $p_{_F}$ (in GeV) 
and $\chi^2_{min}/{d.o.f.}$ for the fixed input values of $m_{sp}=0,~0.15$ GeV 
and $m_q=m_c=1.4,~1.5,~1.6,~1.7$ GeV,  which are the values commonly
used in experimental analyses. As can be noticed, 
these results are  strongly dependent on the input value 
of $m_c$: if we use smaller $m_c$, the best fit value of $p_{_F}$ 
increases, and {\it vise versa}.
In Fig. 1, we also show the theoretical ACCMM model spectrums with
$p_{_F}=0.44,~0.51,~0.59$ GeV (with $m_c=1.5$ GeV, $m_{sp}=0.0$ GeV),
corresponding to dashed-, full-, dotted-line, respectively.
The experimental data and the theoretical predictions are all 
normalized to the semileptonic branching
ratio, ${\cal{BR}}(B \rightarrow X_c l \nu) = 10.49~ \%$,
following the result of CLEO \cite{cleo-exp}.
Previously, we extracted similarly $p_{_F}$ by comparing the theoretical 
prediction with the experimental spectrum of ARGUS \cite{argus2}, 
and we obtained \cite{hkn2} $p_{_F} = 0.27 \pm ^{0.22}_{0.27}$ GeV
for the fixed input values of $m_c=1.5$ GeV and $m_{sp}=0.15$ GeV.
As can be seen from Table I, 
if we fix $m_c=1.5$ GeV and $m_{sp}=0.15$ GeV, then we obtain from the new CLEO
spectrum \cite{cleo-exp}
$p_{_F} = 0.55 \pm ^{0.09}_{0.07}$ GeV  with the minimum $\chi^2$ being about 1.0.
We note that two results are apart each other within one $\sigma$ standard deviation,
but the new result from CLEO has much smaller uncertainties.
In Sections 3 and 4, we give in detail the related physics of this unexpected 
large value of the parameter $p_{_F}$.
\\

\noindent
{\bf \large 3. Average Kinetic Energy of Heavy Quark inside Heavy
Meson}\\

Recently considerable progresses have been achieved on the
relation of the ACCMM model with QCD \cite{bigi4,csaki,ball12}.
Especially
Bigi $et$ $al.$ \cite{bigi4} derived an inequality between the
expectation value of the kinetic energy of the heavy quark
inside the hadron and that of the chromomagnetic operator,
which gives\footnote{This theoretical lower bound could be 
significantly weakened, as shown in \cite{kapustin}, with inclusion 
of the $\alpha_s$ corrections as well as $1/M_Q$ corrections.}
\begin{equation}
\langle {\bf p}^2 \rangle \,\,\ge\,\, {3\over 4}\,\, ({M_V}^2-{M_P}^2)~.
\label{ad1}
\end{equation}
The experimental value of the right hand side of Eq. (\ref{ad1})
is 0.36 ${\rm GeV}^2$ for $B$-meson system \cite{pdg}.
This bound corresponds to $p_{_F}\ge 0.49$ GeV for $B$-meson,
because 
in the ACCMM model the average kinetic energy, 
$\langle {\bf p}^2 \rangle$,
can be calculated from 
\begin{equation}
\langle {\bf p}^2 \rangle =\int d{\bf p}\, p^2 \phi ({\bf p}; p_{_F})\, 
=\, {3\over 2}\,\, {p_{_F}}^2~.
\label{ad1a}
\end{equation}
This relation (\ref{ad1a}) was obtained solely from the fact that
the Gaussian momentum probability distribution was taken in the ACCMM model, 
and therefore the lower bound $p_{_F}\ge 0.49$ GeV is independent of any 
other input parameter values of the ACCMM model, and is much larger than 
the commonly used value $p_{_F} \sim 0.3$ GeV.
Ball $et$ $al.$ \cite{ball12} also calculated 
$\langle {\bf p}^2 \rangle$ 
using the QCD sum rule approach, and obtained
$\langle {\bf p}^2 \rangle = 0.50 \pm 0.10$ ${\rm GeV}^2$
for $B$-meson, corresponding to $p_{_F}=0.58 \pm 0.06$ GeV
from Eq. (\ref{ad1a}).
We note that the heavy quark inside the hadron
possesses more kinetic energy than the value one might expect naively
from the nonrelativistic consideration.
We also note that the Fermi momentum parameter $p_{_F}$ of the ACCMM model is
not a truly free parameter, but is closely related to the average kinetic
energy of heavy quark, which is theoretically calculable in principle.

We consider the relativistic potential model with the
quantum mechanical variational technique to theoretically calculate the average
kinetic energy of $b$-quark inside $B$-meson, and to compare the 
results with the predictions of the HQET.
The potential model has been successful to describe the physics of
$\psi$ and $\Upsilon$ families with the nonrelativistic Hamiltonian
\cite{eich,quigg}.
However, for $B$-meson it has been difficult to apply
the nonrelativistic potential model because of the relativistic motion 
of the light quark inside $B$-meson.
In this work, we  study $B$-meson system with a realistic Hamiltonian,
which is relativistic for the 
light quark and nonrelativistic for the heavy quark,
and adopt the variational method to solve it.
We take the Gaussian function as the trial wave function, and obtain
the ground state energy and wave function by minimizing the 
expectation value of the Hamiltonian. 

For the $B$-meson system we start with the Hamiltonian
\begin{equation}
H=M+{{{\bf p}^2}\over {2M}}+{\sqrt{{\bf p}^2+m^2}}+V(r)~,
\label{f8}
\end{equation}
where $M\equiv m_b$ is the heavy quark mass and 
$m\equiv m_{sp}$ is the $u$- or $d$-quark mass (which corresponds to
the spectator light quark mass in the ACCMM model).
We apply the variational method to the Hamiltonian (\ref{f8})
with the trial wave function
\begin{equation}
\psi ({\bf r})=({{\mu}\over {\sqrt{\pi}}})^{3/2}e^{-{\mu}^2{\bf r}^2/2}~,
\label{f9}
\end{equation}
where the parameter $\mu$ is a variational parameter.
The Fourier transform of $\psi ({\bf r})$ gives the momentum space
wave function $\chi ({\bf p})$, which is also Gaussian,
\begin{equation}
\chi ({\bf p}) = { 1\over ({\sqrt{\pi }}{\mu })^{3/2} }
e^{-{\bf p}^2/2{\mu}^2}~.
\label{f5}
\end{equation}
We note here that the Gaussian momentum probability distribution
of the ACCMM model equals
$\phi({\bf p};p_{_F})=4 \pi |\chi({\bf p};\mu)|^2$.
See Eqs. (\ref{f3}) and (\ref{f5}).
The ground state is given by minimizing the expectation value of $H$,
\begin{equation}
\langle H\rangle =\langle\psi\vert H\vert\psi\rangle =E(\mu )~,
\ \ \
{{d}\over {d\mu }}E(\mu )=0\ \ {\rm{at}}\ \ \mu ={\bar{\mu}}~,
\label{f10}
\end{equation}
and then the value $\bar E \equiv E({\bar{\mu}})$ approximates 
the $B$-meson mass $M_B$,
and at the same time we get ${\bar{\mu}} \equiv p_{_F}$, 
the Fermi momentum parameter in the ACCMM model.
As is well known,
the value of $\bar{\mu}$ or $p_{_F}$ corresponds to the
measure of the radius of the two body bound state, as can be seen from
the relation,
$\langle r\rangle =2/({\sqrt{\pi}}\, {\bar{\mu}} )$
or $\langle r^2{\rangle}^{1/2} =3/(2\, {\bar{\mu}} )$.

We now take in Eq. (\ref{f8}) the Cornell potential,
which is composed of the Coulomb and linear potentials
with a constant term,
\begin{equation}
V(r)=-{{{\alpha}_c}\over {r}}+Kr+V_0 
 \equiv -{4 \over 3}{{{\alpha}_s}\over {r}}+Kr+V_0~.
\label{f13}
\end{equation}
The additive constant $V_0$, which is related to the regularization
concerned with the linear confining potential \cite{lucha1},
is usually known as flavor dependent: $V_0=0$ for heavy-heavy meson system,
$V_0=-0.2$ GeV for $B$-meson system \cite{fulcher}.
We use the value of $K=0.19$ ${\rm GeV}^2$ \cite{hagi} 
for the string tension, and for the parameter
${\alpha}_c\ (\equiv {{4}\over {3}}{\alpha}_s)$
we will consider two values $\alpha_s=0.35$ and $0.24$ separately.
The first choice $\alpha_s=0.35$ is the value which has been determined by
the best fit of $(c{\bar{c}})$ and $(b{\bar{b}})$ bound state spectra
\cite{hagi}, and $\alpha_s=0.24$ is that given by the running coupling
constant for the QCD scale at $M_B$.

With the Gaussian trial wave functions, (\ref{f9}) and (\ref{f5}),
the expectation value of each term of
the Hamiltonian (\ref{f8}) is given as follows:
\begin{eqnarray}
\langle {{\bf p}^2\over 2M}\rangle &=& 
\langle \chi ({\bf p}\,) | {{\bf p}^2\over 2M}|
\chi ({\bf p}\,) \rangle = {3 \over 4M} \mu^2~,
\nonumber\\
\langle \sqrt{{\bf p}\,^2+ m^2} \rangle &=& \langle \chi ({\bf p}\,) 
| \sqrt{{\bf p}\,^2+ m^2} | \chi ({\bf p}\,) \rangle 
= {4\mu \over \sqrt\pi} \int_0^\infty e^{-x^2} \sqrt{x^2 + (m/\mu)^2}
\; x^2dx~,
\nonumber\\
\langle V(r) \rangle &=& \langle \psi ({\bf r}) | -{\alpha_c \over r}
+ Kr +V_0\
|\psi ({\bf r}) \rangle 
= {2 \over \sqrt\pi} (-\alpha_c\mu + {K / \mu} )+V_0~.
\label{f22}
\end{eqnarray}
Then we have
\begin{eqnarray}
&E&({\mu})=\langle H\rangle
\label{f21s}\\
&=& M
+{1\over 2M} \Bigl({3\over 2} \mu^2 \Bigr)
+{2 \over \sqrt\pi} (-\alpha_c\mu + {K / \mu} ) + V_0 +
{4\mu \over \sqrt\pi} \int_0^\infty e^{-x^2} \sqrt{x^2 + (m/\mu)^2}
\; x^2dx~.
\nonumber
\end{eqnarray}
In our previous study \cite{hkn}, we obtained the last integral 
in Eq. (\ref{f21s}) as a power series of $(m/\mu)^2$. And when we write up to
the order of $(m/\mu)^4$, we now get
\begin{eqnarray}
E({\mu}) &=& M 
+{3 \over 4M} \mu^2
+ {2 \over \sqrt\pi} ( -\alpha_c \mu + K/\mu ) + V_0
\nonumber\\
& & + {2\mu \over \sqrt\pi} \biggl[ 1 + {1\over 2} (m/\mu)^2 +
\Bigl({5\over 32} - 2c_1 \Bigr) (m/\mu)^4 + {1\over 4} (m/\mu)^4 \ln(m/\mu)
\biggr] 
\label{af30}\\
& & + {\cal O}\Bigl( (m/\mu)^6 \Bigr)~,
\nonumber
\end{eqnarray}
where $c_1\simeq -0.0975$.
Up to the order of $(m/\mu)^2$, $E(\mu )$ becomes 
\begin{equation}
E(\mu )= M
+{3 \over 4M} \mu^2
+ {2 \over \sqrt\pi} 
\Bigl( (1-\alpha_c )\mu + (K+{1\over 2}m^2)/\mu \Bigr) + V_0~,
\label{af30a}
\end{equation}
and the next order terms $({\cal{O}}((m/\mu)^4))$ contribute only 
less than 1 $\%$. Then, we find the minimum value of $E(\mu )$ in
(\ref{af30a}) by the variational method, and the minimum point is given by
\begin{equation}
{\partial \over \partial\mu }E(\mu )
={3\over 2M}\mu +{2 \over \sqrt\pi}(\beta -\gamma/{\mu}^2)=0~,
\label{z1}
\end{equation}
where
\begin{equation}
\beta \equiv 1-\alpha_c = 1-{4 \over 3}\alpha_s~,\ \ \ \ {\rm and}\ \ \ \
\gamma \equiv K+{1\over 2}m^2~.
\label{z1aa}
\end{equation}
We rewrite Eq. (\ref{z1}) as
\begin{equation}
(\beta {\mu}^2-\gamma )+{b\over M}{\mu}^3=0~,
\label{zb2}
\end{equation}
where $b=3{\sqrt{\pi}}/4$ is a constant.
Then, we expand ${\bar{\mu}}$, which satisfies Eq. (\ref{zb2}), 
as a power series of $1/M$,
\begin{equation}
\bar{\mu} =a_0+a_1{1\over M}+a_2{1\over M^2}+\cdots~,
\label{zb3}
\end{equation}
and by matching the order by the order in (\ref{zb2}), we get
\begin{equation}
a_0=\sqrt{{\gamma\over\beta }},\,\,\,\,\,
a_1=-{b\over 2}\Bigl( {\gamma\over {\beta}^2}\Bigr),\,\,\,\,\,
a_2={5b^2\over 8}\sqrt{{\gamma\over\beta }}
\Bigl( {\gamma\over {\beta}^3}\Bigr)~,\,\,\,\,\,\cdots.
\label{zb4}
\end{equation}
As can be easily seen, since $b/M << 1$, Eq. (\ref{zb2}) has an
approximate solution ${\bar{\mu}}\simeq {\sqrt{\gamma /\beta}} = a_0$.

Using Eqs. (\ref{zb3}) and (\ref{zb4}), we can obtain the numerical
values of the coefficients $a_0$, $a_1$, $a_2$,
and that of $\bar{\mu}$ which minimizes $E(\mu)$ in Eq. (\ref{af30a}),
for $\alpha_s=0.35$ and $0.24$ separately.
We also considered three different values of the light quark mass
$m~(\equiv m_{sp}) = 0.00, ~0.15, ~0.30$ GeV,
in order to see the dependence of the results on the light quark mass $m$.
As we can see from (\ref{z1}) and (\ref{z1aa}), 
the effect of $m$ comes in only through the little modification of
$\gamma$,  because $\gamma \equiv K + m^2/2 \approx K$.
The results of this calculation for $a_0,~a_1,~a_2$ and $\bar \mu$ 
with the input values of $\alpha_s$ and the light quark mass 
$m~(\equiv m_{sp})$ are presented in Table II. 
As previously explained, we fixed\footnote{The numerical value of 
$\bar{\mu}$ is fairly insensitive to the potential we choose.
In Ref. \cite{dsh}, $\bar{\mu}$ has been calculated numerically
from six different potential models, and found to be 
$\bar{\mu}=0.56\pm 0.02$ GeV, where the error
is only the statistical error of the six different results.} 
$K=0.19$ GeV$^2$ and $V_0=-0.2$ GeV.
However, the exact value of $V_0$ is irrelevant in our calculations of 
$\bar{\mu}$, (\ref{zb3}) and (\ref{zb4}),
but it is necessary for the calculation of $\bar{\Lambda}$ 
in Eq. (\ref{zb6}) below.

With $\bar{\mu}$ of (\ref{zb3}) and (\ref{zb4}),
we can get the following expectation values
of the terms in the Hamiltonian (\ref{f8}):
\begin{eqnarray}
\frac{T}{2M}\equiv
\frac{\langle {\bf p}^2 \rangle (\bar{\mu})}{2M} &=&
\frac{3 \bar{\mu}^2}{4M}
\nonumber\\
&=&
{1\over 2M}\left[{3\gamma \over 2\beta }
-{3b\over 2}\sqrt{{\gamma\over\beta }}
\Bigl( {\gamma\over {\beta}^2}\Bigr) {1\over M}
+{9b^2\over 4}\Bigl( {\gamma^2\over {\beta}^4}\Bigr) {1\over M^2}\right]
+{\cal O}({1\over M^3})~,
\label{zb5}
\end{eqnarray}
\begin{eqnarray}
{\bar{\Lambda}}\equiv
\langle \sqrt{{\bf p}^2+m^2} &+& V(r) \rangle (\bar{\mu} ) =
{2\over {\sqrt{\pi}}}(\beta \bar{\mu} +{\gamma\over \bar{\mu}} ) + V_0
\nonumber\\
&=&
\Bigl( V_0 + 2\sqrt{\gamma\beta }\Bigr) +
0 \times {1\over M} +
{b^2\over 4}
\sqrt{{\gamma\over\beta }} \Bigl( {\gamma\over {\beta}^2}\Bigr)
{1\over M^2} + {\cal O}({1\over M^3})~,
\label{zb6}
\end{eqnarray}
Finally, $E(\bar \mu)$ in (\ref{af30a}) is expressed as a power series
in $1/M$,
\begin{eqnarray}
E(\bar \mu) &=& M + {\bar{\Lambda}} + {T \over 2M}
\label{z4b}\\
&\equiv& M+\Bigl( V_0+2\sqrt{\gamma\beta }\Bigr) +
{1\over 2M} \Bigl( {3\over 2}{\gamma\over\beta }\Bigr) 
- \Bigl( {b(3-b)\over 4}
\sqrt{{\gamma\over\beta }} \Bigl( {\gamma\over {\beta}^2}\Bigr)
\Bigr) {1\over M^2}
+{\cal O}({1\over M^3})~.
\nonumber
\end{eqnarray}

In Eq. (\ref{z4b}), the $M$-independent terms come from
$\langle{\sqrt{{\bf p}^2+m^2}}+V(r)\rangle$, which can be considered as
the contributions from the light degrees of freedom.
The term of the order of $1/M$ is from the heavy quark momentum
squared $\langle{\bf p}^2\rangle$, that is, from the average kinetic
energy of the heavy quark inside the heavy-light meson.
Both $\langle{\sqrt{{\bf p}^2+m^2}}+V(r)\rangle$ and 
$\langle{\bf p}^2\rangle$ contribute to the term of the order of $1/M^2$.
In the HQET, the mass of a heavy-light meson is represented \cite{neubert} by
\begin{equation}
M_M=M+{\bar{\Lambda}}+{1\over 2M}(T+{\nu}_{_M}\Omega )
+{\cal O}({1\over M^2})~,
\label{y1aa}
\end{equation}
where
${\bar{\Lambda}}\equiv\lim_{M\rightarrow \infty}(M_M-M)$
is the contribution from the light degrees of freedom, for which
Neubert  obtained \cite{neubert} ${\bar{\Lambda}}=0.57\pm 0.07$ GeV.
$T\equiv \langle{\bf p}^2\rangle$ is the expectation value of
the kinetic energy of the heavy quark (up to $2M$) inside a heavy-light meson,
and $\Omega$ is the expectation value of the energy due to the 
chromomagnetic hyperfine interaction 
with ${\nu}_{_V}=1/4$ and ${\nu}_{_P}=-3/4$.
In this paper we do not consider
the chromomagnetic hyperfine interaction term.
We will present a detailed study on the correspondences
between the relativistic quark model and the heavy quark
effective theory in another forthcoming papar \cite{hknlater}.
Here we calculated only $T$ and ${\bar{\Lambda}}$ up to the order of $1/M^2$
by using (\ref{zb5}) and (\ref{zb6}), and obtained the values shown in
Table III. In Table III, we also show the values of the Fermi momentum
parameter $p_{_F}~(\equiv {\bar{\mu}},~{\rm shown~in~Table~II})$ 
of the ACCMM model using the relation (\ref{ad1a}).

Gremm $et$ $al.$ \cite{gremm} recently
extracted the average kinetic energy, $T \equiv \langle {\bf p}^2 \rangle$,
by comparing the prediction of the HQET \cite{bigi} with
the shape of the inclusive $B \rightarrow X l {\nu}$
lepton energy spectrum \cite{cleo93b} for $E_l \ge 1.5$ GeV, 
in order to avoid the contamination from the secondary leptons of cascade 
decays of $b \rightarrow c \rightarrow s l \nu$.
They obtained
$\lambda_1~(\equiv -T) = -0.35 \pm 0.05\ {\rm GeV}^2$
for $|V_{ub}/V_{cb}|=0.08$ and
$\lambda_1~(\equiv -T) = -0.37 \pm 0.05\ {\rm GeV}^2$
for $|V_{ub}/V_{cb}|=0.1$,
which correspond to
$p_{_F} = 0.48 \pm 0.03$ GeV and
$p_{_F} = 0.50 \pm 0.03$ GeV, repectively.
Their results are remarkably close to the our value in (\ref{tz2})
extracted from the recent model independent lepton energy spectrum of
$B \rightarrow X_c l {\nu}$ \cite{cleo-exp},
as explained in Section 2.

We summarize Section 3 by noting that the value of the Fermi momentum 
parameter of the ACCMM model is $p_{_F}=0.5 \sim 0.6$ GeV and is much 
larger than $\sim 0.3 $ GeV, as can be seen from Table III, and
the heavy quark inside the hadron possesses much more kinetic energy 
than the value one might expect naively
from the nonrelativistic consideration.
\\


\noindent
{\bf \large 4. Dependence of $|V_{cb}|$ and $|V_{ub}/V_{cb}|$
on the Average Kinetic Energy of Heavy Quark inside $B$-meson}\\

Now we consider the dependence on the average kinetic energy 
of $b$-quark
(or equivalently Fermi momentum parameter $p_{_F}$ of the ACCMM model) 
in the $B$-meson semileptonic decay, 
$\langle{\bf p}^2 \rangle$, of the measurements of 
$|V_{cb}|$ and $|V_{ub}/V_{cb}|$.
The $B$-meson inclusive branching fraction is related to the CKM martix
$V_{cb}$ and $V_{ub}$ by
\begin{equation}
{\cal{BR}}(B \rightarrow X l \nu) / \tau_B =
\widetilde{\Gamma}_c |V_{cb}|^2 + \widetilde{\Gamma}_u |V_{ub}|^2 \approx 
\widetilde{\Gamma}_c |V_{cb}|^2~,
\end{equation}
where the factors
$\widetilde{\Gamma}_q   
\equiv {\widetilde{\Gamma}_B} (B\rightarrow X_ql\nu)(p_{_F})$ must be
calculated from theory. (See Eq. (\ref{f1}).)
CLEO has extracted $|V_{cb}| = 0.040 \pm 0.001 \pm 0.004$ 
from their measurements \cite{cleo-exp} of
\begin{eqnarray}
{\cal{BR}}(B \rightarrow X l \nu) 
&=& (10.49 \pm 0.17 \pm 0.43)~ \%~, \nonumber \\
\tau_B &=&  (1.61 \pm 0.04) ~{\rm psec}~,
\end{eqnarray}
and by assuming $\widetilde{\Gamma}_c=(39 \pm 8)$ psec$^{-1}$.
If we instead theoretically calculate $\widetilde{\Gamma}_c$ in the ACCMM model
by using $p_{_F}=0.5 \sim 0.6$ GeV, the result of the ACCMM model becomes
\begin{equation}
|V_{cb}| = |V_{cb}|_{\rm cleo} \times 
\sqrt{{\widetilde{\Gamma}_c^{({\rm CLEO})}} \over 
{\widetilde{\Gamma}_c^{(p_{_F}=0.5 \sim 0.6)}}}
\approx |V_{cb}|_{\rm cleo} \times  1.1 
= 0.044 \pm 0.001 \pm 0.004~.
\label{f28}
\end{equation}
We can easily understand this large correction ($\sim 10$ \%) 
in $|V_{cb}|$ due to the change in $p_{_F}$, 
because within ACCMM model from Eqs. (\ref{f1},\ref{f2})
\begin{eqnarray}
\widetilde{\Gamma}_c \propto m_b^5&=&W^5 \approx (m_{_B}^2-2 m_{_B} p_{_F})^{5/2}~,
\nonumber\\
{\rm and~therefore}~~&~&\widetilde{\Gamma}_c^{(p_{_F}=0.3)} /
\widetilde{\Gamma}_c^{(p_{_F}=0.5)} \approx 1.25~.
\label{f29}
\end{eqnarray}

The ACCMM model also provides an inclusive lepton energy spectrum of 
the $B$-meson semileptonic decay to obtain the value of $|V_{ub}/V_{cb}|$. 
The lepton energy spectrum is useful in separating $b \rightarrow u$
transitions from $b \rightarrow c$,
since the end-point region of the spectrum is completely composed of
$b \rightarrow u$ decays. In applying this method one integrates
(\ref{f4}) in the range $2.3~{\rm GeV}<E_l$ at 
the $B$-meson rest frame, where only
$b \rightarrow u$ transitions exist \cite{cleo2}. 
So we theoretically calculate\footnote{We note that the dependences of  
the lepton energy spectrum  on 
perturbative and non-perturbative QCD corrections \cite{kuhn,bigi} 
as well as on the unavoidable specific model parameters 
({\it e.g.} the parameter $p_{_F}$ of the ACCMM model \cite{alta}) are 
strongest at the end-point region of the inclusive $d \Gamma / d E_l$
distribution.
Therefore, Eq. (\ref{zz1}) may have very limited validity 
for the determination of $|V_{ub}/V_{cb}|$, as shown in \cite{akhoury}.}
\begin{equation}
{\widetilde{\Gamma}}(p_{_F}) \equiv
\int_{2.3} dE_l \,\,
{{d\widetilde{\Gamma}_B}\over {dE_l}}(p_{_F}, m_{sp}, m_q, m_{_B})~.
\label{zz1}
\end{equation}
In (\ref{zz1}) we specified only $p_{_F}$ dependence explicitly in the 
left-hand side.
Then one compares the theoretically calculated 
${\widetilde{\Gamma}}(p_{_F})$ with the experimentally measured width
${\widetilde{\Gamma}}_{exp}$ in the region $2.3~{\rm GeV}<E_l$,
to extract the value of $|V_{ub}|$ from the relation
\begin{equation}
{\widetilde{\Gamma}}_{exp}~=~|V_{ub}|^2~\times {\widetilde{\Gamma}}(p_{_F})~.
\label{z2}
\end{equation}
In the real experimental situations \cite{cleo,argus,argus2,cleo2},
the only measured quantity is the number of events in this region of 
high $E_l$ compared to the total semileptonic events number, 
{\it i.e.} the branching-fraction 
${\widetilde{\Gamma}}_{exp} / {\widetilde{\Gamma}}^{total}_{s.l.}$.
Since the value ${\widetilde{\Gamma}}^{total}_{s.l.}$ is
proportional to $|V_{cb}|^2$, only the combination $|V_{ub} / V_{cb}|^2$ is
extracted.

We now consider the possible dependence of $|V_{ub} / V_{cb}|^2$ 
as a function of the
parameter $p_{_F}$ from the following relation
\begin{equation}
\frac{{\widetilde{\Gamma}}_{exp}} {{\widetilde{\Gamma}}^{total}_{s.l.}}~
\propto~
\left| \frac{V_{ub}}{V_{cb}} \right|^2_{p_{_F}=p_{_F}} 
         \times {\widetilde{\Gamma}}(p_{_F})
=
\left| \frac{V_{ub}}{V_{cb}} \right|^2_{p_{_F}=0.3} 
         \times {\widetilde{\Gamma}}(p_{_F}=0.3)~,
\label{z3}
\end{equation}
where 
$|V_{ub}/V_{cb}|^2_{p_{_F}=p_{_F}}$ is  
determined with an arbitrary value of the Fermi momentum parameter $p_{_F}$.
In the right-hand side we used $p_{_F}$=0.3 GeV because this value is 
commonly used in the experimental determination of 
$\left| {V_{ub}} / {V_{cb}} \right|$.
Then one can get a relation
\begin{equation}
\left| \frac{V_{ub}}{V_{cb}} \right|_{p_{_F}=p_{_F}}
=
\left| \frac{V_{ub}}{V_{cb}} \right|_{p_{_F}=0.3}
\times \sqrt{
\frac{\widetilde{\Gamma}(0.3)}{\widetilde{\Gamma}(p_{_F})} }~.
\label{z4}
\end{equation}

We numerically calculated theoretical ratio 
$\widetilde{\Gamma}(0.3)/\widetilde{\Gamma}(p_{_F})$
by using (\ref{f4}) and (\ref{zz1}) with $m_{sp}=0.15$ GeV, 
$m_q = m_u =0.15$ GeV,
which are the values commonly used by experimentalists,
and $m_{_B}=5.28$ GeV.
We show the values of $|V_{ub}(p_{_F})/V_{ub}(p_{_F}=0.3)|$ as 
a function of $p_{_F}$ in Fig. 2. 
If we use $p_{_F}=0.5 \sim 0.6$ GeV, 
instead of $p_{_F}=0.3$ GeV, in the experimental analysis of 
the end-point region of lepton energy spectrum, 
the value of $|V_{ub}/V_{cb}|$ becomes significantly changed.

Previously the CLEO \cite{cleo2} analyzed with
$p_{_F}=0.3$ GeV the end-point lepton energy spectrum to get
\begin{eqnarray}
10 \times |V_{ub}/V_{cb}|
&=&0.76\pm 0.08 ~~({\rm ACCMM~with}~p_{_F}=0.3~ \cite{cleo2})~,
\nonumber\\
&=&1.01\pm 0.10 ~~({\rm Isgur}~et~al.~{\rm (ISGW)} \cite{isgw})~.
\label{g3}
\end{eqnarray} 
As can be seen, those values differ by two standard deviations\footnote{
There now exists an improved version of ISGW model, 
so-called ISGW2 \cite{scora}, which gives a considerably harder end-point
spectrum than that of ISGW. Therefore, it seems clear that the prediction of ISGW
on $|V_{ub}/V_{cb}|$, Eq. (\ref{g3}), will decrease when re-analyzed by
experimentalists, even though the changes would be small \cite{scora}.}.
However, if we use $p_{_F}=0.5 \sim 0.6$ GeV, the result of the ACCMM model
becomes 
\begin{equation}
10 \times |V_{ub}/V_{cb}|
\approx 1.07 \pm 0.11 ~~({\rm ACCMM~with}~p_{_F}=0.5 \sim 0.6)~,
\end{equation}
and  these two models are in a good
agreement for the value of $|V_{ub}/V_{cb}|$.

We note here that the dependence of  $|V_{ub}/V_{cb}|$ on the parameter
$p_{_F}$ is much stronger compared to that of $|V_{cb}|$. This is because
the $p_{_F}$ dependence of the inclusive distribution 
$d \Gamma / d E_l$ is particularly sensitive if we restrict ourselves 
only in the limited region of end-point, as shown in Eq. (\ref{zz1}).
We would like to emphasize again that the measurements of the
hadronic invariant mass spectrum \cite{kim,cskim} in the inclusive 
$B \rightarrow X_{c(u)} l \nu$ decays can be much more
useful in extracting $|V_{ub}|$ with better theoretical understandings,
where we can use almost the whole region of decay spectrum:
{\it i.e.} in the forthcoming asymmetric $B$-experiments with microvertex 
detectors, BABAR and BELLE, the total separation of $b \rightarrow u$ 
semileptonic decays from the dominant 
$b \rightarrow c$ semileptonic decays would be experimentally viable
using the measurement of inclusive hadronic invariant mass distributions. 
And we could determine the ratio of CKM matrix elements $|V_{ub}/V_{cb}|$  
from the ratio of those measured total integrated decay rates \cite{cskim}, 
which is theoretically described by the phase space factor
and the well-known perturbative QCD correction only.
\\ 


\noindent
{\bf \large 5. Conclusions}\\

The value of the Fermi momentum parameter of the ACCMM model
$p_{_F} \sim 0.3$ GeV, which has been commonly 
used in experimental analyses, has no theoretical or experimental
clear justification.
Therefore, it is strongly recommended to determine the value of
$p_{_F}$ more reliably and independently, 
when we think of the importance of its role in experimental analyses.
It is particularly important in the determination of the value of
$|V_{ub}/V_{cb}|$. We note that the dependence of  $|V_{ub}/V_{cb}|$ 
on the parameter $p_{_F}$ is very strong, because
the inclusive lepton energy distribution 
is particularly sensitive to the variation of $p_{_F}$ 
if we restrict ourselves only in the limited region of end-point. A better 
determination of $p_{_F}$ is also interesting theoretically since it
has its own physical correspondence related to the average kinetic
energy 
$\langle {\bf p}^2 \rangle$ of the heavy quark inside $B$-meson.
Within the ACCMM model the average kinetic energy
is calculated as $\langle {\bf p}^2 \rangle = {3 \over 2} {p_{_F}}^2$,
solely from the fact that the Gaussian momentum probability distribution 
has been taken in the ACCMM model. 
Therefore,  the Fermi momentum parameter $p_{_F}$ of the ACCMM model is
not a truly free parameter, but is closely related to the average kinetic
energy of heavy quark, which is theoretically calculable in principle.

In this context we 
theoretically calculated the value of $p_{_F}$ in the relativistic 
quark model using quantum mechanical variational method.
It turns out that $p_{_F}=0.5 \sim 0.6$ GeV, which is consistent with
the value of $p_{_F}$ determined by comparing the ACCMM model prediction
and the model independent lepton energy spectrum of the CLEO measurement,
$p_{_F}=0.54 \pm ^{0.16}_{0.15}$ GeV.
We note that the value of the Fermi momentum 
parameter of the ACCMM model is much larger than $\sim 0.3$ GeV, and
the heavy quark inside the hadron possesses much more kinetic energy 
than the value one might expect naively from the nonrelativistic consideration.
We also found the correspondences between the relativistic quark
model and the heavy quark effective theory by the $1/M_Q$ expansion,
and the result shows that they are consistent with each other.

If we use $p_{_F}=0.5 \sim 0.6$ GeV, 
instead of $p_{_F}=0.3$ GeV, in the experimental analysis of 
the end-point region of lepton energy spectrum, 
the value of $|V_{ub}/V_{cb}|$ is increased by the factor of 
$1.3 \sim 1.5$ compared with the case of $p_{_F}=0.3$ GeV.
Here we would like to emphasize  that the measurements of the
hadronic invariant mass spectrum in the inclusive 
$B \rightarrow X_{c(u)} l \nu$ decays can be much more
useful in extracting $|V_{ub}|$ with better theoretical understandings.
In future asymmetric $B$ factories with vertex detector, 
the hadronic invariant mass spectrum will offer
alternative ways \cite{kim,cskim} to select $b \rightarrow u$ transitions that 
are much more efficient than selecting the upper end of the lepton energy 
spectrum, with much less theoretical uncertainties.\\


\noindent
{\em Acknowledgements}\\

\indent
The work  was supported 
in part by the Korean Science and Engineering  Foundation, 
Project No. 951-0207-008-2,
in part by Non-Directed-Research-Fund, Korea Research Foundation 1993,
in part by the CTP, Seoul National University, 
in part by Yonsei University Faculty Research Grant,
in part by Daeyang Foundation at Sejong University, 
in part by the Basic Science Research Institute Program,
Ministry of Education 1997,  Project No. BSRI-97-2425, and
in part by COE fellowship of the Japanese Ministry of Education, Science
and Culture.

\pagebreak

\pagebreak

\begin{table}
\begin{center}
\vspace{2cm}
\caption{
The values of $p_{_F}$ (in GeV) and $\chi^2_{min}/{d.o.f}$ for the fixed input 
parameter values $m_{sp}$ and $m_c$ (in GeV).
We derived the values  using $\chi^2$ analysis
by comparing the whole region of experimental electron energy spectrum 
of CLEO [11], which is shown in Fig. 1, 
with the theoretical prediction of ACCMM model, Eq. (4)
using $p_{_F}$ as a free parameter.}
\vspace{2cm}
\begin{tabular}{cccccccccccccr}
& & & & $m_{sp}=$ & 0.00 & & & & $m_{sp}=$ & 0.15 & & &  \\
& & & $m_c=$ 1.4 & 1.5 & 1.6 & 1.7 & &  $m_c=$ 1.4 & 1.5 & 1.6 & 1.7 &  \\
\hline
&&&&&&&&&&&&&\\
& $p_{_F}$ & & 0.64$\pm$0.09 & 0.51$^{+0.08}_{-0.07}$ & 0.40$^{+0.07}_{-0.05}$ &
0.29$^{+0.07}_{-0.06}$ & &  0.69$\pm$0.10 & 0.55$^{+0.09}_{-0.07}$ & 
0.44$^{+0.09}_{-0.06}$ & 0.32$^{+0.08}_{-0.03}$ & & \\
&&&&&&&&&&&&&\\
& $\chi^2_{min}$ & & 1.09 & 1.00 & 1.41 & 2.05 & &  
1.44 & 1.05 & 1.09 & 1.47 & 
\\
\end{tabular}
\end{center}
\end{table}

\pagebreak

\vspace*{0.2cm}

\begin{table}
\begin{center}
\vspace{2cm}
\caption{
The numerical values of 
the coefficients $a_0$, $a_1$, $a_2$ in the $1/M$ expansion of
$\bar{\mu}$, Eq. (20), and the values of
$\bar{\mu}$ which minimizes $E(\mu)$ in Eq. (16). 
We varied $\alpha_s=0.35$, 0.24 and the light quark mass
$m~(\equiv m_{sp})=0.00$, 0.15, 0.30 GeV.}
\vspace{2cm}
\begin{tabular}{ccccccccccccr}
 & &  &$a_0$&$a_1$&$a_2$&${\bar{\mu}}$& \\  \hline
& & $m_{sp}=0.00$&0.60&$-0.60$&1.50&0.54&\\
&$\alpha_s=0.35$& $m_{sp}=0.15$&0.61&$-0.63$&1.62&0.54&\\
& & $m_{sp}=0.30$&0.67&$-0.76$&2.13&0.61&\\  \hline
& & $m_{sp}=0.00$&0.53&$-0.36$&0.63&0.49&\\
&$\alpha_s=0.24$& $m_{sp}=0.15$&0.54&$-0.38$&0.68&0.49&\\
& & $m_{sp}=0.30$&0.59&$-0.46$&0.89&0.54&\\
\end{tabular}
\end{center}
\end{table}

\pagebreak

\vspace*{0.2cm}

\begin{table}
\begin{center}
\vspace{2cm}
\caption{
The average kinetic energy $T$ (up to 2$M$) of the heavy quark,
the contribution of the light degrees of freedom ${\bar{\Lambda}}$,
and the Fermi momentum parameter $p_{_F}$ of $B$-meson system,
for ${\alpha}_s=0.35$, 0.24 and
$m~(\equiv m_{sp})=0.00$, 0.15, 0.30 GeV.
The results obtained by the ${\chi}^2$ analysis of the recent CLEO
lepton energy spectrum, and those from the HQET and the QCD sum rule
approaches are also presented.}
\vspace{2cm}
\begin{tabular}{ccccccccccr}
 & &   &$T$&${\bar{\Lambda }}$&$p_{_F}\, (\equiv {\bar{\mu }})$& \\  \hline
& & $m_{sp}=0.00$ &0.45&0.44&0.54& \\
& $\alpha_s=0.35$& $m_{sp}=0.15$ &0.47&0.46&0.54& \\
& & $m_{sp}=0.30$ &0.57&0.52&0.61& \\  \hline
& & $m_{sp}=0.00$ &0.36&0.52&0.49& \\
& $\alpha_s=0.24$& $m_{sp}=0.15$ &0.38&0.54&0.49& \\
& & $m_{sp}=0.30$ &0.45&0.61&0.54& \\  \hline
& from CLEO data \cite{cleo-exp} & &---&---&0.54$\pm_{0.15}^{0.16}$& \\
& Bigi $et$ $al.$ \cite{bigi4} & &$\geq 0.36$&---&$\geq 0.49$& \\
& Ball $et$ $al.$ \cite{ball12} & &0.50$\pm 0.10$&---&0.58$\pm 0.06$& \\
& Neubert \cite{neubert} & &---&0.57$\pm 0.07$&---& \\
& Gremm $et$ $al.$ \cite{gremm} & $|V_{ub}/V_{cb}|=0.08$ & 0.35$\pm 0.05$ &---&
0.48$\pm 0.03$& \\
& Gremm $et$ $al.$ \cite{gremm} & $|V_{ub}/V_{cb}|=0.10$ & 0.37$\pm 0.05$ &---&
0.50$\pm 0.03$& \\
\end{tabular}
\end{center}
\end{table}

\pagebreak


\epsfxsize=\hsize \epsfysize=0.7\vsize \epsfbox{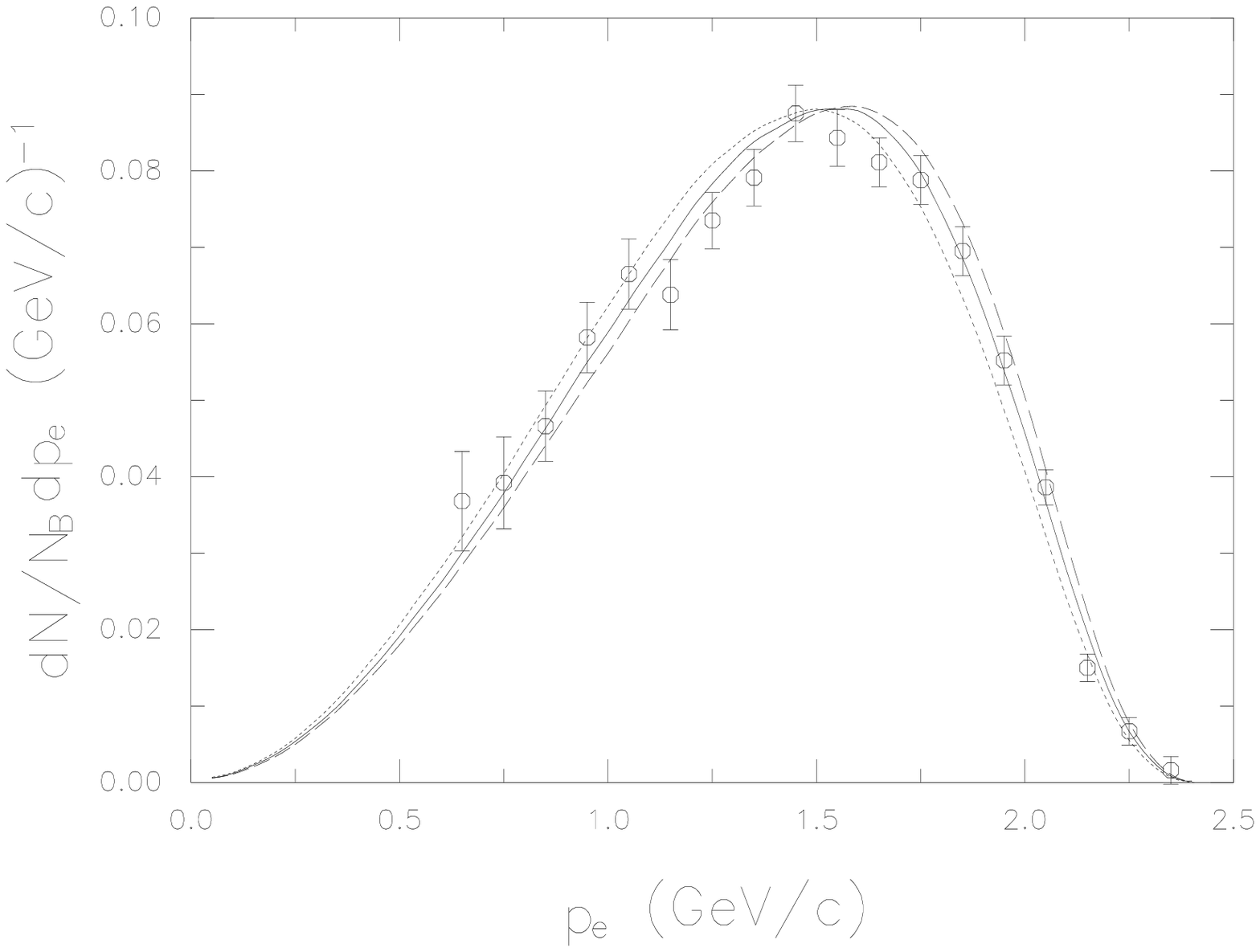}

\noindent
Fig. 1 The normalized lepton energy spectrum of
$B \rightarrow X_c l \nu$
for the whole region of electron energy from the recent
CLEO measurement \cite{cleo-exp}. Also shown are
the theoretical ACCMM model predictions, Eq. (\ref{f4}), using 
$p_{_F}=0.44,~0.51,~0.59$ GeV, corresponding to dashed-, full-,
dotted-line, respectively.
The minimum $\chi^2$  equals to 1.00 with $p_{_F} = 0.51$ GeV.
We fixed $m_{sp}=0.0$ GeV and $m_q=m_c=1.5$ GeV.

\pagebreak

\epsfxsize=\hsize \epsfysize=0.7\vsize \epsfbox{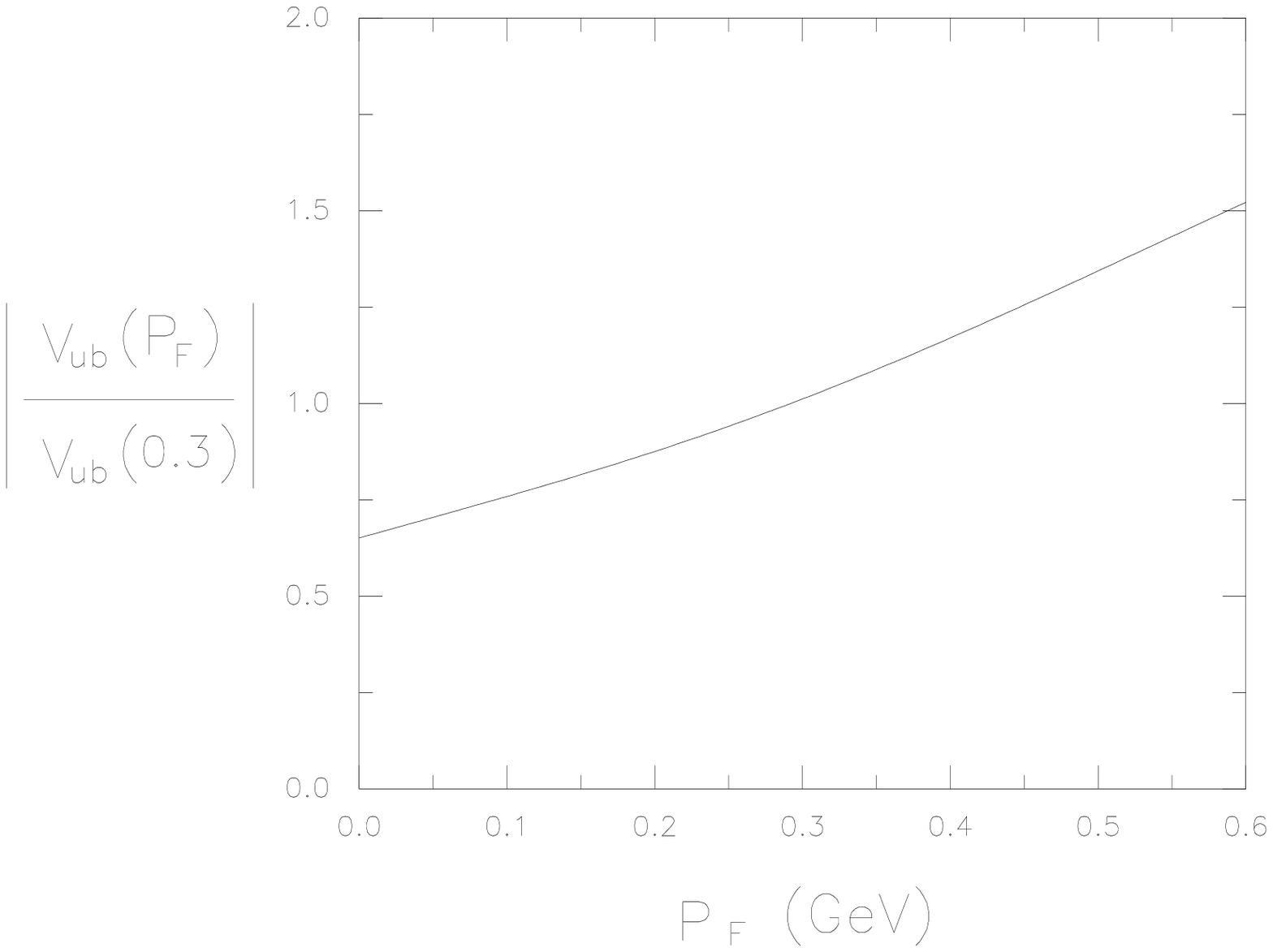}

\noindent
Fig. 2 The ratio  $|V_{ub}(p_{_F})/V_{ub}(p_{_F}=0.3)|$ as 
a function of $p_{_F}$.

\end{document}